# Two-Point Attack on the Two Nonorthogonal State QKD Protocol over a Fiber-Optic Channel

Li Yang[*]

State Key Laboratory of Information Security (Graduate School of Chinese Academy of Sciences),

Beijing 100039, P. R. China

Ling-An Wu

Institute of Physics, Chinese Academy of Sciences, Beijing 100084, P. R. China

The simplest quantum key distribution (QKD) protocol is based on the transmission of two nonorthogonal states [1]. Bennett's original scheme is a multi-photon interferometric scheme using two nonorthogonal low-intensity coherent states, though peoples usually adopt the scheme that employing the single-photon polarization as a carrier of quantum signals [2, 3]. Up to now, several experiments of interferometric scheme have been done, but the bright reference pulses of the original scheme are omitted, this simplification makes the scheme more insecure.

Generally speaking, there are three types of attacks on a secure communication: 1) attack on the algorithm; 2) attack on the protocol; 3) attack on the system. So far thorough analysis of the security of quantum cryptography scheme usually related with the first or the second type of attacks[4,5]. Here we present another kind of eavesdropping strategy: two-point eavesdropping strategy, to attack the two nonorthogonal state QKD system. It is easy to find that the two almost unchangeable parameters, the group velocity of light pulse in fiber and the loss of fiber, render the two nonorthogonal state protocol (single-photon type) over a fiber-optic channel insecure if only the distance between Alice and Bob is longer than a definite length.

Consider that there are two eavesdroppers: Eve1 and Eve2. They can communicate with each other through a classical channel of signal velocity $c$. Eve1 is at the point near Alice's security domain, and Eve2 is at the point near Bob's security domain. Eve1 intercepts the quantum channel and measures every quantum signal. The serious weakness of two-nonorthogonal protocol is that Eve has a non-zero probability of measuring the state of quantum signals exactly. Suppose the carrier particle, transmitted by Alice in one of the two states $|\Psi_1\rangle$ and $|\Psi_2\rangle$, is measured by Bob in one of the two orthonormal bases $\{|\Psi_1\rangle, |\overline{\Psi}_1\rangle\}$ and $\{|\Psi_2\rangle, |\overline{\Psi}_2\rangle\}$, chosen at random. Suppose $|\langle\Psi_1|\overline{\Psi}_2\rangle|^2 = |\langle\Psi_2|\overline{\Psi}_1\rangle|^2 = \sin^2\varphi$, then Eve1's probability of successful measurements is $\frac{1}{2}\sin^2\varphi$. Eves' two-point strategy is rather simple: Eve1 sends a sequence consists of 0, 1, and 2 to Eve2 via the classical channel, where 1 and 2 represent that her measurements are successful in those positions of time: 1 represents the corresponding quantum signal is definitely in the state $|\Psi_1\rangle$, and 2 represents the corresponding quantum signal is definitely in the state $|\Psi_2\rangle$. If only the system is sufficiently long, Eve1 and Eve2 can choose their positions on the channel that satisfy that the loss of the fiber between them is lager than the probability of Eve1's inconclusive outcomes. For example, let us investigate the simplified single-photon B92 protocol. Suppose Eve1's single-photon detector is an ideal one. The probability of Eve1's inconclusive outcomes is $\frac{3}{4}$, that corresponding a loss of $6\,dB$, then the length of fiber between Eve1 and Eve2 should be about 30 kilometers.

---
[*] E-mail: yangli@gscas.ac.cn

Eve2's task is resenting quantum signals $|\Psi_1\rangle$ and $|\Psi_2\rangle$ in accordance with the sequence coming from Eve1. Suppose that all of them, include Alice, Bob, and Eves, use ideal single-photon sources and detectors, it is obvious that Alice and Bob cannot find the two-point attack at all. If we take account of the practical parameters of the sources and detectors, the distance between the two Eves should be longer than that deduced from the ideal case. Besides, Eve2 must use a source of sub-Poissonian statistics if Bob counts the number of photons in every pulse.

The original multi-photon realization of the two nonorthogonal states protocol [1] is designed so ingenious that it can resist not only the coherent amplification attack but also the two-point attack. It is especially interesting that this multi-photon protocol, even realized with weak pulse source, is still rather secure against the beam splitting attack, contrast to BB84 protocol with weak pulse source. In the multi-photon B92 protocol, Eve can only get limited information (less than 10% if each dim signal pulse after Eve1's interferometer contains 0.1 photon averagely) of the raw key regardless of the length of the system, and Alice and Bob can get secure key after privacy amplification. It is clear that the original Bennett 92 protocol is the most secure protocol for fiber-optic QKD system today in the communication wavelength since there is still no stable single-photon source in this wavelength now.